\documentclass{article}

\usepackage{arxiv}

\usepackage[utf8]{inputenc} 
\usepackage[T1]{fontenc}    
\usepackage{hyperref}       
\usepackage{url}            
\usepackage{booktabs}       
\usepackage{amsfonts}       
\usepackage{nicefrac}       
\usepackage{microtype}      
\usepackage{lipsum}
\usepackage{graphicx}
\usepackage{hyperref}
\usepackage{amsfonts}
\usepackage{amsmath}
\usepackage{amssymb}
\usepackage{amsthm}
\usepackage{natbib}
\usepackage{mathtools}
\usepackage{float}
\graphicspath{ {./images/} }

\title{An Integer Linear Program for Periodic Scheduling in Universities}

\author{
 Sina Moradi \\
  Etraab Holding \\
  Economic Expert \\
  Urmia, Iran\\
  \texttt{sina\_moradi1993@yahoo.com} \\
}

\begin{document}
\maketitle
\begin{abstract}
Efficient scheduling of periodic meetings is a critical challenge in various service-oriented domains, including academic settings, healthcare, and legal consultancy. This study presents a robust Integer Linear Programming (ILP) model to optimize the scheduling of faculty-student meetings. The proposed model incorporates practical constraints such as minimum intervals between consecutive meetings, differing time requirements for undergraduate, master’s, and PhD students, and dedicated emergency time slots for unplanned visits. The objective function aims to achieve an equitable distribution of meetings throughout the planning period while prioritizing earlier time slots and seamlessly integrating emergency appointments. To validate the effectiveness of the model, both numerical examples and a case study are examined. The results highlight the model's ability to generate optimal schedules within a computationally efficient framework, leveraging the power of Gurobi optimization software. The model demonstrates significant versatility, extending its applicability beyond academic settings to any scenario requiring periodic scheduling of services, such as patient visits in healthcare or client consultations in legal practices. Future extensions of the model may include dynamic scheduling to adapt to real-time changes in availability and the coordination of schedules across multiple service providers.
\end{abstract}


\section{Introduction}

Scheduling problems are a fundamental area of research in operations and resource management, with roots extending back to the mid-20th century~\citep{giffler1960algorithms}. Among these, periodic scheduling problems have gained particular attention for their relevance to recurring service scenarios, where clients require regular access to service providers~\citep{bar2002minimizing}. Examples include periodic patient visits to healthcare professionals~\citep{gupta2008appointment} or recurring student meetings with academic advisors~\citep{diallo2024optimizing}.

Academic administration, specifically managing the scheduling of faculty meetings within educational institutions is an important in the growth and development of the universities across the glob. Efficient scheduling is crucial in fostering collaborative environments and enhancing productivity among faculty, ultimately contributing to the strategic goals of the institution. This paper focuses on the development of a mathematical model to address the periodic scheduling of faculty-student meetings. Faculty members, as service providers, often dedicate portions of their time to recurring consultations with students, who, in this context, act as service recipients. Scheduling these meetings presents several challenges, such as ensuring equitable distribution across available time periods, accommodating differing needs based on student levels, and reserving slots for unforeseen or emergency consultations.

The proposed model formulates this problem as an Integer Linear Program (ILP), incorporating practical constraints such as minimum gaps between consecutive visits, the varying time requirements of different student categories, and the inclusion of emergency slots. The objective function emphasizes balanced distribution of appointments, prioritization of earlier time slots, and streamlined integration of emergency consultations into the schedule. Integer Linear Programming has been widely adopted for solving scheduling and resource allocation problems due to its versatility and computational efficiency~\citep{geoffrion1972integer,moradi2022efficient,MORADI2024110730}. Recent research underscores the importance of leveraging advanced optimization and multiparadigm frameworks to address complex, interdisciplinary problems, such as cyber-physical security~\citep{Aftabi09122024}, transportation and routing~~\citep{MORADI2023109552,vehicles6030066,MORADI2024110730,MORADI2025125183}, multi-actor system dynamics in information security~\citep{AFTABI2025125681}, or public health systems during pandemics~\citep{nazari14033}.

This research offers a versatile framework applicable not only to academic settings but also to other service domains requiring periodic scheduling. For instance, the framework can be adapted to patient scheduling in healthcare or client scheduling in legal consultancy~\cite{pinedo2015scheduling}. By leveraging the efficiency of ILP solvers like Gurobi~\citep{gurobi}, the proposed approach ensures computational feasibility even for large-scale, real-world scenarios. Additionally, the results demonstrate how the model achieves an optimal balance between operational efficiency and practical scheduling requirements, offering insights for future applications and enhancements.

\section{Methodology}

Considering that the primary objective of this work is to develop a mathematical model for scheduling students' meetings with professors, this research adopts a descriptive-applied approach, aiming to expand practical knowledge in the area of academic scheduling. Faculty members allocate part of their daily work time to student consultations, typically held on a periodic basis. During the planning period (e.g., a month or half an academic year), each student is required to have a specified number of meetings with their advisor. Each meeting occupies one or more half-hour time slots from the professor’s available schedule.

In this framework, each student is treated as a customer, while faculty members correspond to service providers. The professor’s daily schedule is divided into $P$ half-hour time slots, represented by the matrix $F_{jk}$, which identifies the availability of these slots. Students are categorized into three groups: undergraduate, master’s, and PhD. Each group has distinct requirements for meeting durations and frequencies:
\begin{itemize}
    \item Undergraduate students require $R_x$ consecutive half-hour time intervals per meeting and must schedule $T_x$ meetings during the planning period, maintaining a minimum gap of $B_x$ days between successive meetings.
    \item Master’s students require $R_y$ consecutive half-hour time intervals per meeting and $T_y$ meetings, with a minimum gap of $B_y$ days.
    \item PhD students require $R_z$ consecutive half-hour time intervals per meeting and $T_z$ meetings, with a minimum gap of $B_z$ days.
\end{itemize}
In addition, each workday allocates specific time slots for unplanned or emergency visits, ensuring flexibility for addressing unforeseen situations. These emergency slots are scheduled towards the end of each day’s planned meetings, utilizing the first available time intervals after the completion of all scheduled appointments. The objective of this model is twofold:
\begin{enumerate}
    \item Even distribution of meetings: Schedule students' visits as evenly as possible throughout the planning period. For instance, if a student requires four meetings, they should ideally occur on different days, rather than being clustered on a single day.
    \item Optimal placement of emergency slots: Ensure that emergency visits are placed at the end of scheduled meetings to minimize disruptions while efficiently utilizing the professor’s available time.
\end{enumerate}

\subsection{Model Formulation}
In this subsection, we present a mathematical model designed to address the proposed scheduling problem. The model includes the definition of indices, parameters, and decision variables, followed by a detailed discussion of its structure and underlying concepts. The indices, parameters, and decision variables used in the proposed mathematical model are summarized in Tables \ref{tab:indices}, \ref{tab:parameters}, and \ref{tab:decision_variables}, respectively, providing a clear representation of the notations and their corresponding descriptions for ease of understanding. The model is built on the following assumptions:
\begin{itemize}
    \item Each type $s$ student (undergraduate, master’s, or PhD, denoted as $s\in\{x,y,z\}$ must meet with the professor $T_s$ times during the planning period.
    \item During each meeting, a student of type $s$ requires $R_s$ consecutive time slots.
    \item For each student type $s$, there must be a minimum of $B_s$ days between any two consecutive meetings.
    \item Each workday reserves a designated number of time slots to accommodate emergency student visits.
\end{itemize}
These assumptions serve as the foundation for developing the optimization model, ensuring fairness, efficiency, and adaptability in scheduling while accounting for planned and unplanned meetings.

\begin{table}[h!]
\centering
\caption{Indices used in the model.}
\label{tab:indices}
\begin{tabular}{ll}
\hline
\textbf{Index} & \textbf{Description} \\ \hline
$i$            & Student index        \\ 
$j$            & Index of visiting day \\ 
$k$            & Index of available visiting (half-hour) time interval \\ \hline
\end{tabular}
\end{table}

\begin{table}[!ht]
\centering
\caption{Parameters used in the model}
\label{tab:parameters}
\resizebox{\textwidth}{!}{%
\begin{tabular}{ll}
\hline
\textbf{Parameter} & \textbf{Description} \\ \hline
$N_x$             & Total number of undergraduate students \\ 
$N_y$             & Total number of master students \\ 
$N_z$             & Total number of Ph.D. students \\ 
$D$               & Total number of workdays \\ 
$P$               & Total number of available visiting time intervals in each workday \\ 
$T_x$             & Total number of visits for the undergraduate students in the planning period \\ 
$T_y$             & Total number of visits for the master students in the planning period \\ 
$T_z$             & Total number of visits for the Ph.D. students in the planning period \\ 
$B_x$             & Minimum gap in days between visits of undergraduate students \\ 
$B_y$             & Minimum gap in days between visits of master students \\ 
$B_z$             & Minimum gap in days between visits of Ph.D. students \\ 
$R_x$             & Number of consecutive time intervals required for undergraduate students per visit \\ 
$R_y$             & Number of consecutive time intervals required for master students per visit \\ 
$R_z$             & Number of consecutive time intervals required for Ph.D. students per visit \\ 
$L$               & Number of available time intervals for emergency visits per day \\ 
$F_{jk}$         & Matrix indicating if the professor is available on day $j$ at time interval $k$ \\ \hline
\end{tabular}
}
\end{table}

\begin{table}[!ht]
\centering
\caption{Decision variables used in the model}
\label{tab:decision_variables}
\resizebox{\textwidth}{!}{%
\begin{tabular}{ll}
\hline
\textbf{Decision Variable} & \textbf{Description} \\ \hline
$X_{ijk}$ & Equals 1 if undergraduate student $i$ visits on day $j$ at time interval $k$, otherwise 0. \\ 
$Y_{ijk}$ & Equals 1 if master student $i$ visits on day $j$ at time interval $k$, otherwise 0. \\ 
$Z_{ijk}$ & Equals 1 if Ph.D. student $i$ visits on day $j$ at time interval $k$, otherwise 0. \\ 
$E_{jk}$   & Equals 1 if time interval $k$ on day $j$ is assigned to emergency visits, otherwise 0. \\ 
$W_{ij}$   & Equals 1 if student $i$ visits on day $j$, otherwise 0. \\ \hline
\end{tabular}
}
\end{table}

The constraints of the model are as follows:
\begin{equation}
\sum_{k=1}^P E_{jk} = L, \quad \forall j \in \{1, \dots, D\}
\label{eq:constraint1}
\end{equation}

\begin{equation}
\sum_{i=1}^{N_x} X_{ijk} + \sum_{i=1}^{N_y} Y_{ijk} + \sum_{i=1}^{N_z} Z_{ijk} + E_{jk} \leq F_{jk}, 
\quad \forall j \in \{1, \dots, D\}, \forall k \in \{1, \dots, P\}
\label{eq:constraint2}
\end{equation}

\begin{equation}
\sum_{k=1}^P X_{ijk} = R_x \cdot W_{ij}, \quad \forall i \in \{1, \dots, N_x\}, \forall j \in \{1, \dots, D\}
\label{eq:constraint3}
\end{equation}

\begin{equation}
X_{ijk} + X_{ijh} \leq 1, \quad \forall i \in \{1, \dots, N_x\}, \forall j \in \{1, \dots, D\}, \forall k \in \{1, \dots, P\},  k+R_x \leq h \leq P
\label{eq:constraint4}
\end{equation}

\begin{equation}
\sum_{k=1}^P Y_{ijk} = R_y \cdot W_{i,j}, \quad \forall i \in \{1, \dots, N_y\}, \forall j \in \{1, \dots, D\}
\label{eq:constraint5}
\end{equation}

\begin{equation}
Y_{ijk} + Y_{ijh} \leq 1, \quad \forall i \in \{1, \dots, N_y\}, \forall j \in \{1, \dots, D\}, \forall k \in \{1, \dots, P\},  k+R_y \leq h \leq P
\label{eq:constraint6}
\end{equation}

\begin{equation}
\sum_{k=1}^P Z_{ijk} = R_z \cdot W_{ij}, \quad \forall i \in \{1, \dots, N_z\}, \forall j \in \{1, \dots, D\}
\label{eq:constraint7}
\end{equation}

\begin{equation}
Z_{ijk} + Z_{ijh} \leq 1, \quad \forall i \in \{1, \dots, N_z\}, \forall j \in \{1, \dots, D\}, \forall k \in \{1, \dots, P\},  k+R_z \leq h \leq P
\label{eq:constraint8}
\end{equation}

\begin{equation}
\sum_{j=1}^D \sum_{k=1}^P X_{ijk} = R_x \cdot T_x, \quad \forall i \in \{1, \dots, N_x\}
\label{eq:constraint9}
\end{equation}

\begin{equation}
\sum_{j=1}^D \sum_{k=1}^P Y_{ijk} = R_y \cdot T_y, \quad \forall i \in \{1, \dots, N_y\}
\label{eq:constraint10}
\end{equation}

\begin{equation}
\sum_{j=1}^D \sum_{k=1}^P Z_{ijk} = R_z \cdot T_z, \quad \forall i \in \{1, \dots, N_z\}
\label{eq:constraint11}
\end{equation}

\begin{equation}
\sum_{k=1}^P \sum_{j=t}^{t+B_x} X_{ijk} \leq R_x, \quad \forall i \in \{1, \dots, N_x\}, \forall t \in \{1, \dots, D-B_x\}
\label{eq:constraint12}
\end{equation}

\begin{equation}
\sum_{k=1}^P \sum_{j=t}^{t+B_y} Y_{ijk} \leq R_y, \quad \forall i \in \{1, \dots, N_y\}, \forall t \in \{1, \dots, D-B_y\}
\label{eq:constraint13}
\end{equation}

\begin{equation}
\sum_{k=1}^P \sum_{j=t}^{t+B_z} Z_{ijk} \leq R_z, \quad \forall i \in \{1, \dots, N_z\}, \forall t \in \{1, \dots, D-B_z\}
\label{eq:constraint14}
\end{equation}

Constraint \eqref{eq:constraint1} governs the allocation of emergency time slots for each day. It ensures that exactly $L$ time slots are reserved for emergency visits on every working day. This allocation provides flexibility to handle unplanned student meetings while adhering to the time constraints.

Constraint \eqref{eq:constraint2} is linked to the matrix $F_{jk}$, which indicates the professor's availability. If the professor is present and has free time during a specific time slot, $F_{jk}$ equals 1; otherwise, it equals 0. The constraint ensures two critical conditions: 1) Only time slots during which the professor is available are assigned to students or emergency visits; 2) No more than one entity (undergraduate, master’s, Ph.D. student, or emergency visit) is assigned to a single time slot.

Constraints \eqref{eq:constraint3} and \eqref{eq:constraint4} define the allocation rules for undergraduate students. In constraint \eqref{eq:constraint3}, each undergraduate student must be assigned $R_x$ consecutive time slots for their visit on a specific day. In constraint \eqref{eq:constraint4}, the allocated $R_x$ time slots must be consecutive. The same rules apply to master’s and Ph.D. students, as detailed in Constraints \eqref{eq:constraint5}, \eqref{eq:constraint6}, \eqref{eq:constraint7}, and \eqref{eq:constraint8}.

Constraint \eqref{eq:constraint9} ensures that the total required time slots for undergraduate students are allocated over the planning period (e.g., a semester). Each undergraduate student is required to meet their professor $T_x$ times during this period, with each visit requiring $R_x$ time slots. Thus, the total time slots allocated to each undergraduate student during the planning period must equal $T_x \times R_x$. Constraints \eqref{eq:constraint10} and \eqref{eq:constraint11} extend the logic of Constraint \eqref{eq:constraint9} to master’s and Ph.D. students. They ensure that the total required time slots for these student groups are allocated appropriately over the planning period.

Constraint \eqref{eq:constraint12} enforces a minimum gap of $B_x$ days between two consecutive visits for undergraduate students. It ensures that during this interval, the total number of allocated time slots does not exceed the number of slots required for a single visit by the undergraduate student. This prevents overlapping or overly frequent visits within the specified minimum gap period. Similar to Constraint \eqref{eq:constraint12}, constraints \eqref{eq:constraint13} and \eqref{eq:constraint14} apply to master’s and Ph.D. students, ensuring that the required minimum gap between consecutive visits ($B_y$ for master’s and $B_z$ for Ph.D. students) is maintained.

The objective function is defined as
\begin{equation}
\text{Minimize } 
2 \cdot \left( \sum_{j=1}^D \sum_{k=1}^P k \cdot 
\left( \sum_{i=1}^{N_x} X_{ijk} + \sum_{i=1}^{N_y} Y_{ijk} + \sum_{i=1}^{N_z} Z_{ijk} \right) \right) 
+ \sum_{j=1}^D \sum_{k=1}^P k \cdot E_{jk}
\label{eq:objective}
\end{equation}

The objective function in \eqref{eq:objective} reflects three key goals:

\begin{enumerate}
    \item Uniform distribution of visits: To ensure an even distribution of both scheduled and emergency visits across all working days, the weight assigned to scheduling student meetings is uniform across days.
    \item Preference for early time slots: Scheduled visits are prioritized for earlier hours of the day. For example, if a student can be scheduled in either the first or the fourth half-hour slot, the first slot will be preferred. To achieve this, assigning a visit to time interval $k$ increases the objective function by 2, pushing assignments toward earlier slots.
    \item Emergency visit placement: Emergency visits are scheduled immediately after the regular meetings of the day. If an emergency visit is allocated to time interval $k$, the objective function increases by $k$, encouraging these visits to occupy the earliest available slots after regular appointments.
\end{enumerate}
By combining these aims, the objective function ensures that the schedule is both efficient and practical, balancing planned and unplanned visits while prioritizing early time slots and maintaining an even distribution of meetings.

The latest advancements in this study include the daily allocation of specific time intervals for emergency student consultations, enabling the accommodation of unplanned visits as they arise. Constraint \eqref{eq:constraint1} ensures the reservation of $L$ emergency time slots each day, maintaining flexibility for unforeseen needs. A notable improvement is the prioritization of student consultations earlier in the day. For instance, if a student has the option to attend a meeting in either the first or fourth half-hour slot, preference is given to the earlier time. This prioritization is implemented in the objective function, which adds a penalty of two for regular appointments scheduled in later intervals, thereby incentivizing earlier scheduling. Emergency appointments are strategically placed immediately following pre-scheduled ones. To achieve this, the objective function increases by the interval number for emergency consultations, encouraging their placement in the earliest available slots after regular meetings. This optimization problem is efficiently solved using the Gurobi solver, leveraging its advanced capabilities to produce practical and effective scheduling solutions.

\section{Computational Experiments and Results}
\subsection{Numerical study}
In this section, the proposed model is validated through a simple numerical example with small dimensions, evaluated using the Gurobi software. The use of a small-scale problem ensures that the solution space is limited, allowing for a thorough examination of all conditions specified by the problem designer. This approach facilitates the verification of the model's adherence to the desired constraints and objectives.

Table \ref{tab:small_example} presents the details of the student visit scheduling problem in small dimensions, providing a clear overview of the input parameters and problem setup.

\begin{table}[h!]
\centering
\caption{Parameters of the small example}
\label{tab:small_example}
\begin{tabular}{|l|c|c|c|c|}
\hline
\textbf{Student Type} & \textbf{R} & \textbf{N} & \textbf{T} & \textbf{B} \\ \hline
Undergraduate         & 1          & 2          & 3          & 1          \\ \hline
Master                & 2          & 3          & 2          & 2          \\ \hline
PhD                   & 3          & 2          & 1          & 3          \\ \hline
\end{tabular}
\end{table}

In this simple example, the planning is conducted over a 7-day period, with each day divided into 6 time slots. It is assumed that the professor is available during all time slots throughout the planning period. The optimal solution, obtained using the Gurobi software, is presented in Table \ref{tab:small_example_results}.

\begin{table}[h!]
\centering
\caption{Results for the small example}
\label{tab:small_example_results}
\begin{tabular}{|l|c|}
\hline
\textbf{Metric}              & \textbf{Value} \\ \hline
Number of constraints         & 5958           \\ \hline
Number of variables           & 1155           \\ \hline
Objective value               & 139            \\ \hline
\end{tabular}
\end{table}

The optimal timetable for this example is presented in Table \ref{tab:optimal_solution_small_example}. The symbols $U_i$, $M_i$, and $P_i$ represent the undergraduate, master’s, and PhD students assigned to a specific time slot, respectively. Additionally, the symbol $E$ indicates that the time slot has been allocated for emergency visits. For instance, the fourth time slot on Wednesday is designated for emergency visits. In this schedule, Master student number 1 is required to visit on the first day (Monday) during time slots 1 and 2. The second visit for this student is scheduled on the fifth day (Friday) during time slots 2 and 3.

\begin{table}[h!]
\centering
\caption{Optimal solution of the small example}
\label{tab:optimal_solution_small_example}
\begin{tabular}{|l|c|c|c|c|c|c|}
\hline
\textbf{Day}     & \textbf{1} & \textbf{2} & \textbf{3} & \textbf{4} & \textbf{5} & \textbf{6} \\ \hline
Monday           & $M_1$         & $M_1$         & $M_3$         & $M_3$         & $E$          &            \\ \hline
Tuesday          & $M_2$         & $M_2$         & $U_1$         & $U_2$         & $E$          &            \\ \hline
Wednesday        & $P_2$         & $P_2$         & $P_2$         & $E$          &            &            \\ \hline
Thursday         & $P_1$         & $P_1$         & $P_1$         & $U_1$         & $E$          &            \\ \hline
Friday           & $U_2$         & $M_1$         & $M_1$         & $E$          &            &            \\ \hline
Monday           & $M_2$         & $M_2$         & $U_1$         & $E$          &            &            \\ \hline
Tuesday          & $U_2$         & $M_3$         & $M_3$         & $E$          &            &            \\ \hline
\end{tabular}
\end{table}

\subsection{Case study}
A case study in this section is designed using the proposed model and based on the schedule of a selected professor from a university (name withheld for anonymity). The case study involves planning and solving the research problem for one semester, equivalent to 16 weeks. Table \ref{tab:weekly_schedule} presents the professor’s weekly schedule, spanning five workdays. Each workday is divided into 22 half-hour intervals.

\begin{table}[h!]
\centering
\caption{The professor’s schedule for one week (five workdays)}
\label{tab:weekly_schedule}
\resizebox{\textwidth}{!}{%
\begin{tabular}{|l|c|c|c|c|c|c|c|c|c|c|c|c|c|c|c|c|c|c|c|c|c|c|}
\hline
\textbf{Day}      & \textbf{1} & \textbf{2} & \textbf{3} & \textbf{4} & \textbf{5} & \textbf{6} & \textbf{7} & \textbf{8} & \textbf{9} & \textbf{10} & \textbf{11} & \textbf{12} & \textbf{13} & \textbf{14} & \textbf{15} & \textbf{16} & \textbf{17} & \textbf{18} & \textbf{19} & \textbf{20} & \textbf{21} & \textbf{22} \\ \hline
Monday            & 1          & 1          & 1          & 1          & 1          & 1          & 1          & 1          & 1          & 1           & 1           & 1           & 1           & 1           & 1           & 1           & 1           & 1           & 1           & 1           & 1           & 1           \\ \hline
Tuesday           & 1          & 1          & 1          & 1          & 0          & 0          & 0          & 0          & 0          & 0           & 0           & 0           & 0           & 0           & 0           & 1           & 1           & 1           & 1           & 1           & 1           & 1           \\ \hline
Wednesday         & 0          & 0          & 0          & 0          & 0          & 0          & 0          & 0          & 0          & 0           & 0           & 0           & 0           & 0           & 0           & 1           & 1           & 1           & 1           & 1           & 1           & 1           \\ \hline
Thursday          & 0          & 0          & 0          & 0          & 0          & 0          & 0          & 0          & 0          & 0           & 0           & 0           & 0           & 0           & 0           & 1           & 1           & 1           & 1           & 1           & 1           & 1           \\ \hline
Friday            & 1          & 1          & 1          & 1          & 1          & 1          & 1          & 0          & 0          & 0           & 0           & 0           & 0           & 0           & 0           & 0           & 0           & 0           & 0           & 0           & 0           & 0           \\ \hline
\end{tabular}%
}
\end{table}

The data associated with this problem includes the number of students at each academic level, the total number of visits required by each student during the planning period, the minimum gap between consecutive meetings, and the number of time intervals required for each type of student per visit. These details are comprehensively presented in Table \ref{tab:case_study_parameters}.

\begin{table}[h!]
\centering
\caption{Parameters of the case study}
\label{tab:case_study_parameters}
\begin{tabular}{|l|c|c|c|c|}
\hline
\textbf{Student Type} & \textbf{R} & \textbf{N} & \textbf{T} & \textbf{B} \\ \hline
Undergraduate         & 1          & 8          & 10         & 1          \\ \hline
Master                & 2          & 6          & 12         & 2          \\ \hline
PhD                   & 3          & 5          & 6          & 3          \\ \hline
\end{tabular}
\end{table}

The solution to the problem, obtained using the Gurobi optimization software, is detailed in Table \ref{tab:case_study_results}. The table provides key metrics, including the total number of constraints, the number of variables, and the objective value achieved for the optimal solution. These results reflect the computational complexity and efficiency of the proposed model, demonstrating Gurobi's capability to handle a large-scale problem while adhering to all specified constraints and objectives. The reported objective value highlights the balanced allocation of time slots for student visits and emergency intervals, aligning with the goals of uniform distribution and prioritization of early scheduling.

\begin{table}[h!]
\centering
\caption{Results for the case study}
\label{tab:case_study_results}
\begin{tabular}{|l|c|}
\hline
\textbf{Metric}              & \textbf{Value} \\ \hline
Number of constraints         & 180545         \\ \hline
Number of variables           & 36080          \\ \hline
Objective value               & 3682           \\ \hline
\end{tabular}
\end{table}

Given that the planning period spans half an academic year (16 weeks), presenting the full schedule for all weeks would result in an overly lengthy text. To address this, Table \ref{tab:professor_schedule_case_study} summarizes the results for week 4, providing a representative view of the scheduling outcomes for a single week. Additionally, Table \ref{tab:master_student_schedule} presents the detailed schedule of visits for the fifth master’s student across all 16 weeks of the semester. This table outlines the assigned days and time slots for each scheduled meeting. For instance, the third meeting for this student is planned on the fifth day of the second week of the semester, during time slots 3 and 4. This example illustrates how the proposed model efficiently accommodates the constraints and requirements of individual students while ensuring an optimal allocation of time slots. These tables highlight the flexibility and precision of the model in handling both weekly schedules and semester-long planning for individual students.

\begin{table}[h!]
\centering
\caption{Professor’s schedule for one week (five workdays)}
\label{tab:professor_schedule_case_study}
\resizebox{\textwidth}{!}{%
\begin{tabular}{|l|c|c|c|c|c|c|c|c|c|c|c|c|c|c|c|c|c|c|c|c|c|c|}
\hline
\textbf{Day}      & \textbf{1} & \textbf{2} & \textbf{3} & \textbf{4} & \textbf{5} & \textbf{6} & \textbf{7} & \textbf{8} & \textbf{9} & \textbf{10} & \textbf{11} & \textbf{12} & \textbf{13} & \textbf{14} & \textbf{15} & \textbf{16} & \textbf{17} & \textbf{18} & \textbf{19} & \textbf{20} & \textbf{21} & \textbf{22} \\ \hline
Monday            & $P_3$         & $P_3$         & $P_3$         & $U_4$         & $U_2$         & $U_7$         & $U_3$         & $U_1$         & $M_6$         & $M_6$          & $E$           &             &             &             &             &             &             &             &             &             &             &             \\ \hline
Tuesday           & $M_2$         & $M_2$         & $U_8$         & $U_6$         & *          & *          & *          & *          & *          & *           & *           & *           & *           & *           & $E$           &             &             &             &             &             &             \\ \hline
Wednesday         & *          & *          & *          & *          & *          & *          & *          & *          & *          & *           & *           & *           & *           & *           & $E$           &             &             &             &             &             &             \\ \hline
Thursday          & *          & *          & *          & *          & *          & *          & *          & *          & *          & *           & *           & *           & *           & *           & $E$           &             &             &             &             &             &             \\ \hline
Friday            & $M_3$         & $M_3$         & $U_3$         & $U_4$         & $U_6$         & $U_1$         & $E$          &             &             &             &             &             &             &             &             &             &             &             &             &             &             \\ \hline
\end{tabular}%
}
\end{table}

\begin{table}[h!]
\centering
\caption{The schedule of visits of the fifth master student for 16 weeks}
\label{tab:master_student_schedule}
\begin{tabular}{|l|c|c|c|c|c|c|c|c|c|c|c|c|}
\hline
\textbf{Week} & 1  & 2  & 2  & 3  & 5  & 5  & 7  & 11 & 12 & 13 & 15 & 16 \\ \hline
\textbf{Day}  & 1  & 1  & 5  & 5  & 2  & 5  & 2  & 1  & 1  & 1  & 2  & 1  \\ \hline
\textbf{Slot} & 9,10 & 6,7 & 3,4 & 3,4 & 1,2 & 2,3 & 1,2 & 1,2 & 4,5 & 7,8 & 1,2 & 6,7 \\ \hline
\end{tabular}
\end{table}

\section{Discussion \& Conclusion}
The results of solving the numerical examples demonstrate that the careful definition of constraints and the objective function is both effective and leads to desired outcomes across various scheduling problems. A notable strength of the proposed model lies in its formulation as an integer linear programming (ILP) problem. This approach ensures computational efficiency, resulting in short solution times even for large, real-world problems.

The versatility of the presented model extends beyond the scheduling of professors’ appointments. It is applicable to a wide range of service organizations where clients require periodic services. Examples include periodic client consultations with lawyers and regular patient visits to doctors. In such contexts, the model can accommodate different categories of clients or patients, each with unique time requirements for their appointments. Additionally, the inclusion of daily emergency intervals provides the flexibility to address unplanned referrals, further enhancing the model's practical utility.

In the case of patient scheduling, the model could be tailored to handle diverse patient needs by categorizing them based on required time intervals and incorporating emergency slots for urgent visits. This adaptability highlights the potential of the proposed framework in addressing complex scheduling challenges in other service-based industries.

Looking ahead, future research could expand the scheduling model to accommodate scenarios involving multiple service providers, such as professors, lawyers, or doctors, who share common clients or students. This would involve developing a system that retains the model's existing features while optimizing the scheduling of appointments across multiple providers. Such an extension would ensure conflict-free schedules, taking into account the availability of all providers and shared responsibilities. Moreover, the integration of dynamic scheduling elements, such as real-time adjustments to address sudden changes in availability or client needs, could further enhance the applicability of the model. Leveraging advanced computational methods or incorporating machine learning techniques to predict client demands and optimize schedules dynamically could also be explored.

\bibliographystyle{plainnat}  
\bibliography{references}  

\end{document}